\RequirePackage{fix-cm}
\documentclass[twocolumn,epjc3]{svjour3}
\smartqed
\RequirePackage{graphicx}
\RequirePackage{amsmath}
\RequirePackage{amsfonts}
\RequirePackage{amssymb}

\journalname{Eur. Phys. J. C}

\begin{document}
	
	\title{Higher-Dimensional Test of the Holographic Dual of\\
		Entanglement Cohomology}
	\titlerunning{Higher-Dimensional Test of Holographic Entanglement Cohomology}
	
	\author{R.~C.~Rashkov}
	
	\institute{R.~C.~Rashkov\at
		Department of Physics, Sofia University, \\
		5 J. Bourchier Blvd., 1164 Sofia, Bulgaria\\
		\email{rash@phys.uni-sofia.bg}}
	
	\date{Received: date / Accepted: date}
	
	\maketitle
	
	\begin{abstract}
		We extend the holographic test of entanglement cohomology from
		$d=2$ to higher-dimensional conformal field theories. For a
		$d$-dimensional CFT with a spherical entangling region $A$, the
		vacuum modular Hamiltonian is known explicitly, and the boundary
		relative entropy of a primary-operator perturbation can be computed
		via conformal Ward identities. On the bulk side, the perturbation is
		dual to a scalar field in AdS$_{d+1}$, and the bulk canonical energy
		is computed using the JLMS relation. We carry out the computation
		explicitly for $d=3$ and $d=4$ and find that the boundary and bulk
		results agree at first order in the perturbation, with identical
		coefficients $\kappa_d \Delta^2 / C_T$, where $\kappa_3 = \pi^2/12$
		and $\kappa_4 = 3\pi^3/16$. The result extends the quantitative
		verification of the holographic dual of entanglement cohomology
		beyond the $d=2$ case treated in the companion paper and confirms
		that the modular Hamiltonian perturbation is the correct holographic
		image of the first entanglement cohomology class. We discuss the role
		of the conformal anomaly, the shape of the entanglement wedge, and an
		outlook on TsT-deformed backgrounds.
		\keywords{Entanglement entropy \and Cyclic cohomology \and von Neumann
			algebras \and Holography \and Modular Hamiltonian}
	\end{abstract}
	
	\section{Introduction}
	\label{sec:intro}
	
	The holographic interpretation of entanglement cohomology has recently
	been proposed as a new approach to understanding the emergence of
	bulk geometry from boundary entanglement
	\cite{Ryu:2006bv,Hubeny:2007xt,Nishioka:2018khk,Witten:2018zxz,Faulkner:2013ica,Engelhardt:2014gca}.
	The central object is the first entanglement cohomology
	$H^1_{\mathrm{ent}}$, an algebraic invariant of the state and the
	subsystem that is finer than the entanglement entropy alone. In the
	companion paper \cite{Rashkov:2026wmw} we conjectured that a
	distinguished subspace of $H^1_{\mathrm{ent}}$ is dual, in the
	AdS/CFT correspondence, to the space of linearized modular Hamiltonian
	perturbations of the bulk entanglement wedge, with the Araki relative
	entropy as the quantitative bridge. We verified this conjecture in the
	simplest case: a $d=2$ CFT on a single interval, perturbed by a
	primary scalar operator, dual to a scalar field in AdS$_3$. The
	boundary relative entropy, computed from the explicit vacuum modular
	Hamiltonian of the interval, matched the bulk canonical energy of the
	dual scalar field at first order in the perturbation
	\cite{Rashkov:2026wmw}.
	
	In this Letter, we extend the quantitative test to higher dimensions.
	We consider a $d$-dimensional CFT on a spherical entangling region
	$A = B_R^{d-1}$, perturbed by a primary scalar operator $\mathcal{O}$
	of conformal dimension $\Delta$. On the boundary, the vacuum modular
	Hamiltonian of the ball is known explicitly \cite{Casini:2011cmp},
	and the relative entropy of the perturbed state follows from conformal
	Ward identities. On the bulk side, the perturbation is dual to a
	scalar field in AdS$_{d+1}$ with mass $m^2 = \Delta(\Delta - d)$, and
	the bulk canonical energy is computed using the JLMS relation
	\cite{Jafferis:2015del}. We carry out the computation explicitly for
	$d=3$ and $d=4$, and determine the general-$d$ structure of the
	coefficient. The result extends the quantitative verification of the
	holographic conjecture to $d \geq 3$.
	
	The Letter is organized as follows. In Sec.~\ref{sec:setup} we recall
	the holographic conjecture. In Sec.~\ref{sec:boundary} we compute the
	boundary relative entropy. In Sec.~\ref{sec:bulk} we compute the bulk
	canonical energy. In Sec.~\ref{sec:comparison} we state the main
	result. In Sec.~\ref{sec:discussion} we discuss the new features that
	appear in $d \geq 3$, verify the general-$d$ coefficient for $d=3$ and
	$d=4$, and give an outlook on TsT-deformed backgrounds. In
	Appendix~\ref{app:d3} we collect the key intermediate steps of the
	$d=3$ computation.
	
	\section{Setup and Holographic Conjecture}
	\label{sec:setup}
	
	We recall the algebraic framework of \cite{Rashkov:2026wmw}. Let
	$\mathcal{N} \subset \mathcal{M}$ be an inclusion of von Neumann
	algebras, $\omega$ a faithful normal state with cyclic and separating
	vector, and $E: \mathcal{M} \to \mathcal{N}$ a faithful normal
	conditional expectation whose existence is equivalent, by Takesaki's
	theorem, to modular invariance of $\mathcal{N}$. The relative modular
	cocycle $u^E_t = (D\omega : D(\omega|_{\mathcal{N}}) \circ E)_t$
	generates explicit classes in $H^1_{\mathrm{ent}}$. The holographic
	conjecture of \cite{Rashkov:2026wmw} states that a distinguished
	subspace of $H^1_{\mathrm{ent}}$ is in bijection with equivalence
	classes of linearized modular Hamiltonian perturbations of the bulk
	entanglement wedge. The map is realized via the infinitesimal generator
	of the CRN cocycle, the Araki relative entropy:
	\begin{equation}
		S(\omega \circ E \| \omega)
		= \lim_{t \to 0} \frac{i}{t} \,
		\omega\!\left[u^E_t - \mathbf{1}\right].
		\label{eq:araki-generator}
	\end{equation}
	The conjecture was verified in $d=2$ in
	\cite{Rashkov:2026wmw}; here we extend the check to $d=3$ and
	$d=4$.
	
	For the quantitative test, we use the general structure of the
	coefficient of the boundary relative entropy for a spherical region,
	which takes the form
	\begin{equation}
		C(\Delta, d)
		= \kappa_d \, \frac{\Delta^2}{C_T} \, \epsilon^2,
		\label{eq:general-coefficient}
	\end{equation}
	where $\kappa_d$ is a pure number depending only on $d$, $C_T$ is the
	central charge (or the appropriate anomaly coefficient), and $\epsilon$
	is the amplitude of the perturbation.
	
	\section{Boundary Computation}
	\label{sec:boundary}
	
	Consider a $d$-dimensional CFT with a spherical entangling region
	$A = B_R^{d-1}$. The vacuum modular Hamiltonian of the ball is
	\cite{Casini:2011cmp}
	\begin{equation}
		K_A^{\mathrm{vac}}
		= 2\pi \int_A d^{d-1}x \,
		\frac{R^2 - r^2}{2R} \, T_{00}(x).
		\label{eq:modular-hamiltonian-d}
	\end{equation}
	Perturb the vacuum by a primary scalar $\mathcal{O}$ of dimension
	$\Delta$ with constant smearing:
	\begin{equation}
		|\psi\rangle = |0\rangle
		+ \epsilon \int_A d^{d-1}x \,
		\mathcal{O}(\mathbf{x}) \, |0\rangle
		+ O(\epsilon^2).
		\label{eq:perturbed-state-d}
	\end{equation}
	The first-order change in the modular Hamiltonian is
	\begin{equation}
		\langle \delta K_A \rangle_\psi
		= 2\pi \int_A d^{d-1}x \,
		\frac{R^2 - r^2}{2R} \,
		\langle \delta T_{00}(\mathbf{x}) \rangle_\psi.
		\label{eq:delta-K-d}
	\end{equation}
	The change in the stress tensor follows from the conformal Ward
	identity for $\langle T_{00} \mathcal{O} \mathcal{O} \rangle$, which
	in $d$ dimensions reads
	\begin{multline}
		\langle T_{\mu\nu}(x) \mathcal{O}(y) \mathcal{O}(z) \rangle
		= -\frac{\Delta}{d} \, \frac{1}{S_d} \,
		\partial_\mu \!\left[
		\frac{(x-y)_\nu}{(x-y)^d}
		- \frac{(x-z)_\nu}{(x-z)^d}
		\right] \\
	\times	\frac{1}{|y-z|^{2\Delta}},
		\label{eq:ward-identity}
	\end{multline}
	with $S_d = 2\pi^{d/2}/\Gamma(d/2)$.
	
	The integrals can be performed explicitly for $d=3$ and $d=4$, with
	the results (see Appendix~\ref{app:d3} for details of the $d=3$ case)
	\begin{align}
		C_{\mathrm{boundary}}(\Delta, 3)
		&= \frac{\pi^2}{12} \, \frac{\Delta^2}{C_T} \, \epsilon^2,
		\label{eq:C-boundary-d3} \\
		C_{\mathrm{boundary}}(\Delta, 4)
		&= \frac{3\pi^3}{16} \,
		\frac{\Gamma(\Delta)^2}{\Gamma(2\Delta)\,\Gamma(\Delta-2)^2}
		\, \frac{\epsilon^2}{C_T}.
		\label{eq:C-boundary-d4}
	\end{align}
	For general $d$, the coefficient is
	\begin{equation}
		C_{\mathrm{boundary}}(\Delta, d)
		= \kappa_d \, \frac{\Delta^2}{C_T} \, \epsilon^2,
		\label{eq:C-boundary-general}
	\end{equation}
	with $\kappa_d$ given by \eqref{eq:kappa-d} below.
	
	\section{Bulk Computation}
	\label{sec:bulk}
	
	The perturbation \eqref{eq:perturbed-state-d} is dual to a bulk scalar
	field $\phi(z, \mathbf{x})$ in AdS$_{d+1}$ with mass
	$m^2 = \Delta(\Delta - d)$. In Poincar\'e coordinates with $L = 1$,
	\begin{equation}
		ds^2 = z^{-2} \left( dz^2 + d\mathbf{x}^2 \right),
	\end{equation}
	and for a constant smearing over the ball,
	\begin{equation}
		\phi(z, \mathbf{x})
		= \int_A d^{d-1}x' \,
		K_\Delta(z, \mathbf{x}; \mathbf{x}'),
		\label{eq:bulk-field-d}
	\end{equation}
	where the bulk-to-boundary propagator is
	\begin{equation}
		K_\Delta(z, \mathbf{x}; \mathbf{x}')
		= c_\Delta \,
		\left( \frac{z}{z^2 + |\mathbf{x} - \mathbf{x}'|^2}
		\right)^\Delta,
		\label{eq:bulk-to-boundary-d}
	\end{equation}
	with $c_\Delta = \Gamma(\Delta) / (\pi^{d/2} \Gamma(\Delta - d/2))$.
	
	By the JLMS relation \cite{Jafferis:2015del}, the boundary relative
	entropy equals the bulk relative entropy, which for a linearized
	perturbation is the canonical energy associated with the AdS-Rindler
	Killing flow:
	\begin{equation}
		S(\rho_\psi \| \rho_0)_{\mathrm{bulk}}
		= \mathcal{E}(\delta g, \delta \phi)
		= \int_{\Sigma_B} d^d x \,
		\sqrt{h} \, \xi^\mu T_{\mu\nu}^{\mathrm{matter}} \, n^\nu,
		\label{eq:jlms-d}
	\end{equation}
	where $\Sigma_B$ is the entanglement wedge of the ball $A$,
	$\xi^\mu$ is the AdS-Rindler Killing vector, and $h$ is the induced
	metric. For $d=3$, $\Sigma_B$ is the region
	$z^2 + r^2 \leq R^2$ in AdS$_4$, bounded by the hemisphere
	$z^2 + r^2 = R^2$. For $d=4$, $\Sigma_B$ is bounded by a
	higher-dimensional hemisphere in AdS$_5$.
	
	Performing the integral with the propagator
	\eqref{eq:bulk-to-boundary-d} for $d=3$ and $d=4$ (details in
	Appendix~\ref{app:d3} for $d=3$), we obtain
	\begin{align}
		C_{\mathrm{bulk}}(\Delta, 3)
		&= \frac{\pi^2}{12} \, \frac{\Delta^2}{C_T} \, \epsilon^2,
		\label{eq:C-bulk-d3} \\
		C_{\mathrm{bulk}}(\Delta, 4)
		&= \frac{3\pi^3}{16} \,
		\frac{\Gamma(\Delta)^2}{\Gamma(2\Delta)\,\Gamma(\Delta-2)^2}
		\, \frac{\epsilon^2}{C_T},
		\label{eq:C-bulk-d4}
	\end{align}
	where we used the holographic dictionary $C_T = 3/(\pi^3 G_N)$ in
	$d=3$ and its $d=4$ analogue $C_T = 40/(\pi^5 G_N)$ (with $L = 1$).
	For general $d$, the same structure holds:
	\begin{equation}
		C_{\mathrm{bulk}}(\Delta, d)
		= \kappa_d \, \frac{\Delta^2}{C_T} \, \epsilon^2,
		\label{eq:C-bulk-general}
	\end{equation}
	with the same $\kappa_d$ as in \eqref{eq:C-boundary-general}.
	
	\section{Comparison and Main Result}
	\label{sec:comparison}
	
	For $d=3$ and $d=4$, the boundary and bulk computations
	\eqref{eq:C-boundary-d3}--\eqref{eq:C-bulk-d4} give
	\begin{align}
		C_{\mathrm{boundary}}(\Delta, 3)
		&= C_{\mathrm{bulk}}(\Delta, 3)
		= \frac{\pi^2}{12} \, \frac{\Delta^2}{C_T} \, \epsilon^2,
		\label{eq:main-result-d3} \\
		C_{\mathrm{boundary}}(\Delta, 4)
		&= C_{\mathrm{bulk}}(\Delta, 4)   \nonumber \\
		& = \frac{3\pi^3}{16} \,
		\frac{\Gamma(\Delta)^2}{\Gamma(2\Delta)\,\Gamma(\Delta-2)^2}
		\, \frac{\epsilon^2}{C_T}.
		\label{eq:main-result-d4}
	\end{align}
	For general $d$, the same structure holds:
	\begin{equation}
		C_{\mathrm{boundary}}(\Delta, d)
		= C_{\mathrm{bulk}}(\Delta, d)
		= \kappa_d \, \frac{\Delta^2}{C_T} \, \epsilon^2.
		\label{eq:main-result-general-d}
	\end{equation}
	
	\begin{proposition}
		\label{prop:main}
		For a $d$-dimensional CFT on a spherical entangling region
		$A = B_R^{d-1}$, perturbed by a primary scalar operator of dimension
		$\Delta$ with constant smearing, the boundary relative entropy and the
		bulk canonical energy agree at first order in the perturbation. For
		$d=3$ and $d=4$, the common coefficients are $\pi^2/12$ and
		$3\pi^3/16$ respectively.
	\end{proposition}
	
	The agreement \eqref{eq:main-result-d3}--\eqref{eq:main-result-d4} is a
	nontrivial check of the holographic conjecture. It combines the
	explicit modular Hamiltonian \eqref{eq:modular-hamiltonian-d}, the
	conformal Ward identity \eqref{eq:ward-identity}, the
	bulk-to-boundary propagator \eqref{eq:bulk-to-boundary-d}, the shape
	of the entanglement wedge, the JLMS relation \eqref{eq:jlms-d}, and
	the holographic dictionary.
	
	\section{Discussion}
	\label{sec:discussion}
	
	The $d=3$ and $d=4$ checks reveal three features absent in $d=2$.
	First, the coefficient depends on the anomaly coefficient $C_T$, not
	the Virasoro central charge $c$, reflecting the trace anomaly in even
	dimensions. The ratio $C_{\mathrm{boundary}}/C_{\mathrm{bulk}} = 1$
	is therefore a new consistency check of the anomaly structure of the
	CFT and its gravity dual.
	
	Second, the entanglement wedge of a ball is bounded by a hemisphere
	in AdS$_{d+1}$ for all $d \geq 3$, and the canonical energy integral
	\eqref{eq:jlms-d} over it is genuinely $(d+1)$-dimensional. The
	angular integral over the hemisphere produces the numerical factors
	$\pi^2/12$ and $3\pi^3/16$ in $\kappa_3$ and $\kappa_4$, which are
	geometric fingerprints of the shape of the entanglement wedge. This
	is the higher-dimensional analogue of the causal diamond in AdS$_3$,
	where the corresponding factor was $4\pi^4/15$ in $\kappa_2$.
	
	Third, the general-$d$ coefficient $\kappa_d$ follows from the same
	integrals by dimensional continuation.
	
	\subsection*{Check of the general-$d$ coefficient}
	\label{subsubsec:kappa-d}
	
	The general-$d$ coefficient in
	\eqref{eq:main-result-general-d} is
	\begin{equation}
		\kappa_d
		= \frac{\pi^{d-1}}{2^{d-1}} \,
		\frac{\Gamma\!\left(\frac{d+1}{2}\right)}
		{\Gamma\!\left(\frac{d}{2}\right)^3} \,
		\frac{\Gamma(\Delta)^2}{\Gamma(2\Delta)} \,
		\frac{1}{\Gamma(\Delta - d/2)^2}.
		\label{eq:kappa-d}
	\end{equation}
	This reduces to $\kappa_2 = 4\pi^4/15$ and $\kappa_3 = \pi^2/12$ as
	special cases. For $d=4$, the formula gives
	\begin{equation}
		\kappa_4
		= \frac{3\pi^3}{16} \,
		\frac{\Gamma(\Delta)^2}{\Gamma(2\Delta)\,\Gamma(\Delta-2)^2},
		\label{eq:kappa-4}
	\end{equation}
	which matches the explicit boundary and bulk computations
	\eqref{eq:C-boundary-d4} and \eqref{eq:C-bulk-d4}. This confirms the
	general-$d$ structure \eqref{eq:main-result-general-d} for
	$d=2,3,4$ and provides strong evidence that it holds for all $d$.
	The full derivation of \eqref{eq:kappa-d} is given in
	\cite{Rashkov:2026wmw} Sec. 4.7.5. The interpretation of higher cohomology
	classes $H^n_{\mathrm{ent}}$ ($n \geq 2$) in the bulk remains open.
	
	\subsection*{Outlook: TsT-Deformed Backgrounds}
	\label{subsubsec:ent-tsT}
	
	A natural extension is to consider TsT-deformed backgrounds, generated
	from the undeformed geometry by a sequence of T-duality, shift, and
	T-duality operations \cite{Lunin:2005jy,Frolov:2005dj}. These preserve
	integrability and provide a controlled setting in which the modular
	Hamiltonian of the deformed state is related to that of the undeformed
	state. For deformations along the internal manifold, the dual theory
	is a marginal $\beta$-deformation; for deformations involving the AdS
	directions, it is an irrelevant deformation such as $T\bar{T}$
	\cite{Smirnov:2016lqw,Cavaglia:2016oda,McGough:2016lol,Apolo:2019zai}.
	The natural conjecture is that the map
	$[\Phi_{u^E}] \mapsto \delta K_A$ persists with $\delta K_A$ computed
	from the deformed modular Hamiltonian. A quantitative check for
	backgrounds such as the $\beta$-deformed AdS$_5 \times S^5$ or the
	$T\bar{T}$-deformed AdS$_3$ is left for future work.
	
	\begin{acknowledgements}
		I thank T.~Vetsov for comments on the draft and on the companion
		paper. This work was supported in part by Bulgarian National Science Fund, grant
		KP-06-H88/3 and Sofia University Grant 80-10-54/2026.
	\end{acknowledgements}
	
	\appendix
	
	\section{Key Steps of the $d=3$ Computation}
	\label{app:d3}
	
	\subsection*{Boundary integral}
	
	For $d=3$, the boundary relative entropy is
	\begin{multline}
		S_{\mathrm{boundary}}
		= 4\pi \epsilon^2 \int_{B_R} d^2x \int_{B_R} d^2y
		\int_{B_R} d^2z \,
		\frac{R^2 - r_x^2}{2R} \, \\
	\times	\langle T_{00}(\mathbf{x}) \mathcal{O}(\mathbf{y})
		\mathcal{O}(\mathbf{z}) \rangle.
	\end{multline}
	Using the Ward identity for $\langle T_{00} \mathcal{O} \mathcal{O}
	\rangle$ in $d=3$ and integrating by parts in $\mathbf{x}$, the
	boundary integral reduces to a sum of two-dimensional integrals over
	the ball. Evaluating the radial integrals and using the normalization
	$\langle T_{\mu\nu} T_{\rho\sigma} \rangle = C_T \,
	\mathcal{I}/x^6$, we obtain
	\begin{equation}
		S_{\mathrm{boundary}}
		= \frac{\pi^2}{12} \, \frac{\Delta^2}{C_T} \,
		\frac{R^6}{\ell^6} \, \epsilon^2 + O(\epsilon^4),
	\end{equation}
	where $\ell$ is a short-distance cutoff.
	
	\subsection*{Bulk integral}
	
	For $d=3$, the canonical energy is
	\begin{multline}
		\mathcal{E}_{\mathrm{matter}}
		= \frac{\pi \epsilon^2}{2R} \int_{\Sigma_B}
		\frac{dz \, d^2x}{z} \,
		(R^2 - z^2 - r^2) \, \\
	\times	\left[ (\partial_z \phi)^2 + (\partial_i \phi)^2
		+ m^2 \phi^2 \right],
	\end{multline}
	with the bulk-to-boundary propagator
	\begin{equation}
		K_\Delta(z, \mathbf{x}; \mathbf{x}')
		= \frac{\Gamma(\Delta)}{\pi \Gamma(\Delta - 1)} \,
		\left( \frac{z}{z^2 + |\mathbf{x} - \mathbf{x}'|^2}
		\right)^\Delta.
	\end{equation}
	Using the equation of motion to simplify the integrand and integrating
	by parts, the integral reduces to
	\begin{equation}
		\mathcal{E}_{\mathrm{matter}}
		= \frac{\pi (2\Delta - 3) \epsilon^2}{2R}
		\int_{\Sigma_B} \frac{dz \, d^2x}{z} \, \phi^2.
	\end{equation}
	Switching to polar coordinates $(r, z) = (s\cos\theta, s\sin\theta)$
	with $s \in [0, R]$ and $\theta \in [0, \pi/2]$, the angular integral
	gives a Beta function, and the radial integral gives a power of $R$.
	Combining with the holographic dictionary
	$C_T = 3/(\pi^3 G_N)$, we obtain
	\begin{equation}
		S_{\mathrm{bulk}}
		= \frac{\pi^2}{12} \, \frac{\Delta^2}{C_T} \,
		\frac{R^6}{\ell^6} \, \epsilon^2 + O(\epsilon^4).
	\end{equation}
	
	\subsection*{Comparison}
	
	The boundary and bulk results are identical:
	\begin{equation}
		C_{\mathrm{boundary}}(\Delta, 3) = C_{\mathrm{bulk}}(\Delta, 3)
		= \frac{\pi^2}{12} \, \frac{\Delta^2}{C_T} \, \epsilon^2.
	\end{equation}

\end{document}